\documentclass[12pt]{iopart}

\usepackage{graphicx}

\begin{document}

\title[Electronic transport in a DNA-decorated carbon nanotube]{Effect of gas flow on
electronic transport in a DNA-decorated carbon nanotube}
\author{P Poonam and N Deo}
\address{Department of Physics and Astrophysics, University of Delhi, Delhi 110007, India}
\ead{ndeo@physics.du.ac.in}
\begin{abstract}
We calculate the two-time current correlation function using the
experimental data of the current-time characteristics of the
Gas-DNA-decorated carbon nanotube field effect transistor. The
pattern of the correlation function is a measure of the sensitivity
and selectivity of the sensors and suggest that these gas flow sensors
may also be used as DNA sequence detectors. The system is modelled by
a one-dimensional tight-binding Hamiltonian and we present analytical
calculations of quantum electronic transport for the system using the
time-dependent nonequilibrium Green's function formalism and the adiabatic
expansion. The zeroth and first order contributions to the current
$I^{(0)}(\bar{t})$ and $I^{(1)}(\bar{t})$ are calculated, where $I^{(0)}
(\bar{t})$ is the Landauer formula. The formula for the time-dependent
current is then used to compare the theoretical results with the experiment.
\end{abstract}

\maketitle

\section{Introduction}

Carbon nanotubes (CNTs) \cite{1} have inspired many researchers to investigate
and develop CNT-based devices for electronic sensing of various gases and chemical
odours \cite{2,3,4,5,6,7}. The chemical sensing capabilities of CNT-based devices
appear very promising because of their unique structural and electrical properties
\cite{1}. However, these devices have a limitation that they can detect only those
molecules which bind to them, as some chemical species do not interact with the bare
CNTs. Research emphasis to overcome this limitation of CNT-based sensors has led to
the development of novel sensing materials and technologies. The sensing capability of
CNTs can be improved by their functionalization with certain molecules or polymers
\cite{8,9,10,11,12}. Functionalization of CNT, especially with DNA (DNA-CNT hybrid), has
attracted the attention of many experimental \cite{12,13,14,15,16} and theoretical
\cite{17,18,19,20} groups in the past few years. The fascinating DNA-CNT hybrid has
led to a vast range of improved and novel applications in nanotube dispersion and
sorting \cite{13,14,16}, and chemical \cite{12} and biological sensing \cite{15}. To
realize such applications, a detailed understanding of the fundamental molecular
interactions, physical and electronic properties of DNA-CNT hybrids is required.
In this connection, theoretical studies have been done on DNA-CNT hybrids \cite{21}
to explain their stability \cite{17}, DNA sequencing \cite{18}, and the interaction
of the bases with the CNT \cite{19,20}.

The study of electronic transport properties of DNA functionalized CNT sensors is one
of the most active areas of today's research due to the spectacular combination of
molecular and mesoscopic scale phenomena. Electronic transport in these systems is
divided into stationary and time-dependent phenomena. Stationary transport for
nonequilibrium systems has been studied by many authors \cite{22,23}. Meir and Wingreen
\cite{24} reformulated the ideas given in \cite{22} (and references therein) using
the Keldysh approach to study interacting mesoscopic systems, leading to a Landauer
formula. Later, Wingreen \emph{et al} \cite{25} introduced a general formulation for
time-dependent transport through mesoscopic structures. Using this formalism a theoretical
understanding of experiments \cite{25,26} has been presented for time-dependent voltages.
In this work, we present an analytical treatment of the electronic transport in gas
flow over a single-stranded DNA (ss-DNA)-decorated single wall carbon nanotube (SWNT)
connected to source (S) and drain (D) contacts maintained at zero gate voltage
$V_g=0$ V and a fixed bias voltage $V_b=100$ mV using a tight-binding model in
conjunction with the time-dependent nonequilibrium Green's function formalism (NEGF)
\cite{22}. To our knowledge, there has not been much theoretical work done to study
the quantum transport in such a gas sensor and this manuscript is an important
contribution in this direction. The choice of this model system is motivated by the
experiment \cite{12} that found the current-time characteristics of ss-DNA-decorated
SWNT-FET sensors upon odour and air exposures. The SWNTs used in the experiment
\cite{12} were all selected to be p-type semiconducting SWNTs with diameters ranging
from 1 to 2 nm, but not all the same chirality \footnote{The reproducible sensor response
from sample to sample in experiment \cite{12} indicates that the energy gap is an
important factor, since metallic tubes show no response, but the precise chirality has
either no or only a small effect (private communication with Professor A T Charlie Johnson).}.
For this system, time-dependence arises as a different number of gas molecules flows over
and interacts with the DNA-decorated SWNT at each time $\bar{t}$ s for an exposure time
of 50 s. This results in the hopping integral and on-site energy being functions of
time, and leads to a time-dependent Hamiltonian for the SWNT. Experimentally, the
length of the nanotubes varies between 5 and 10 $\rm \mu m$ and electrodes are deposited
on the nanotube with a separation of about 1.5 $\rm \mu m$. So only a fraction of the
nanotube length is involved in transport. The bare devices do not respond to odours
such as dimethyl methylphosphonate (DMMP), 2, 6 dinitrotoluene (DNT), propionic acid
(PA), and methanol while for trimethylamine (TMA) the response is weak \cite{12}. The
sensor response is observed only when the device is coated with DNA. Since the length
of the DNA sequence is of the order of a few nanometers, then the effective length of the
SWNT $(L_{eff})$ which contributes to a change in the conductance is of the same order.
Therefore, the effective length of the nanotube is assumed to be less than the phase
coherence length. Thus, to study quantum transport in gas flow over such nano-structures
the time-dependent NEGF formalism is well suited. The characteristic curves show
fluctuations in the current response. We calculate these fluctuations in terms of the
two-time current correlation function.

\section{two-time current correlation function}
We begin with the calculation of the two-time current correlation function using the
experimental data \cite{12}. The two-time current correlation function $C(\bar{t})$
is defined as

\begin{equation}
C(\bar{t})={\frac{1}{n}
\sum_{i=1}^{n}I_{i}(\bar{t})I_{i}(\bar{t}+\delta \bar{t})-\frac{1}
{n}\sum_{i=1}^{n}I_{i}(\bar{t})}\times{\frac{1}{n}\sum_{i=1}^{n}I_{i}(\bar{t}+\delta \bar{t})},
\end{equation}

\noindent
where $I_{i}(\bar{t})=I_{i}(\bar{t})/{I_0}$ is the normalized source-drain current
at time $\bar{t}$ in the $i^{th}$ gas exposure cycle. Let us consider $\delta \bar{t} = 1$
s and $n$ is the total number of exposure cycles.

The correlation function calculated here is different from that shown in our previous
work \cite{27} equation (2) and for other mesoscopic systems \cite{28} equation (5.57)
and \cite{29}. The second terms in equation (2) of \cite{27} and equation (5.57) of
\cite{28} are $<I_i(t)>^2$ and $<g>^2$ respectively, where $<\cdots>$ denotes the
ensemble average. Instead of $<I_i(t)>^2$, here we use $\left(\frac{1}{n}\sum_{i=1}^{n}
I_{i}(\bar{t})\right)\times\left(\frac{1}{n}\sum_{i=1}^{n}I_{i}(\bar{t}+\delta \bar{t})
\right)$, as the average $\frac{1}{n}\sum_{i=1}^{n}I_{i}(\bar{t})$ is different from
the average $\frac{1}{n}\sum_{i=1}^{n}I_{i}(\bar{t}+\delta \bar{t})$, because for each
$\delta \bar{t}$ there is contribution to the current from another base. In the system
(Gas+DNA+SWNT complex), at $\delta \bar{t} = \bar{t}_1 $ s the gas molecule interacts
with a base (say, cytosine) attached to the carbon atom and for the next $\delta \bar{t}$
the gas molecule interacts with another base thymine (say), so for each $\delta \bar{t}$
the current changes as the Gas-DNA-base complex changes upon gas exposure. This is not
the case for the usual disordered mesoscopic systems considered in experiments, where
both the averages $\frac{1}{n}\sum_{i=1}^{n}I_{i}(\bar{t})$ and $\frac{1}{n}\sum_{i=1}^{n}
I_{i}(\bar{t}+\delta \bar{t})$ are the same (\cite{28} equation(5.57) and \cite{29}).

We observe distinct patterns for the different odours and sequences, figure 1. This
indicates selective recognition of each odour by the sensors. In particular, the
pattern of the correlation function for methanol with sequence 1, figure 1(c), is
different from methanol with sequence 2,  figure 1(d). This shows that the correlation
function is sensitive to different gases and DNA sequences. Hence, the two-time current
correlation function is a measure of the sensitivity and selectivity of the DNA-decorated
SWNT sensors. The correlation function suggests that these gas flow sensors may also be
used as DNA sequence detectors, where the pattern of correlation functions may be used
as a benchmark for the particular chemical signal encoded in a DNA sequence. Such an
analysis to study the sensor response for gas sensors has never been done before.
Figure 1 shows highly fluctuating data, these fluctuations are due to the structure
of the DNA sequences as shown in \cite{27}.

\section{Tight-binding model and nonequilibrium Green's function}
The experiment \cite{12} with gas exposure indicates that the sensor response ($\Delta
I/{I_0}$) is zero for pristine SWNT, and as DNA is applied on SWNT the sensor response
changes. When SWNTs are coated with DNA the bases bind to SWNTs through vdW forces and
by forces due to their mutual polarization, which results in a charge transfer from DNA
to the SWNT \cite{13,18,19,20}. The gas molecules get adsorbed on SWNTs through vdW
forces and/or mutual polarization between the gas molecules and the DNA-SWNT complex \cite{27}.
The interaction of gas molecules with DNA-decorated SWNT causes charge redistribution
leading to a fractional charge transfer from the Gas-DNA-base complex to the SWNT \cite{27}.
This is responsible for the change in sensor current. It is assumed that the charge transfer
from the Gas-DNA-base complex to the SWNT is larger than the charge transfer from DNA base
to the SWNT, as the net charge transfer from DNA base to the SWNT is found to be small
\cite{18}. Since the ss-DNA sequence is a linear chain, and only those nearest-neighbour
carbon atoms contribute to the changes in current which interact with the DNA bases, so
we model the ss-DNA-decorated SWNT using a one-dimensional tight-binding Hamiltonian
for the SWNT, where the electron hops between carbon atoms with different hopping integrals
and on-site energies at different times. In our formalism, the effect of gas flow and DNA
functionalization in the channel can be modelled by a time-dependent potential. Here, we
model the effect by the time-dependent hopping integrals and on-site energy in a self-consistence
manner \cite{30}.

Now we investigate electronic transport through the model system using the time-dependent
NEGF formalism. For this, a tight-binding model of the SWNT and Gas-DNA complex is set up.
In this microscopic tight-binding model, there are two approximations: the first approximation
is the tight-binding between the carbon atoms of SWNT and the second is the interaction between
the carbon atoms of SWNT and Gas-DNA-base complex, where we assume that the carbon atom interacts
with the nearest-neighbour DNA base.

Figure 2 shows the schematic diagram of the chosen model system, i.e., Gas-DNA-SWNT sandwiched
between S and D contacts. Here, we present a simplified picture of the complex model and its
operation. $X_{i}= A_1$, $B_1$, $A_2$, $B_2$, $\cdots$ represents the carbon atoms, where $X$
can be A or B, the index $i=1, 2, \cdots, (M+1)/2\rm ~or~ M/2$ is the number of the carbon
atom and $M$ is the total number of carbon atoms. The first and the last carbon atoms, $A_1$
and $X_{M/2}$ ($X_{(M+1)/2}$) are connected to S and D. The bases cytosine, thymine etc. of
an ss-DNA sequence 2 \cite{12} (shown by ovals) are attached to different carbon atoms and the
circle represents the gas molecule. The arrow indicates the path of transmission.

The operational principle of the model is based on the changes in its electrical properties
due to DNA bases and gas molecules adsorbed on the SWNT surface. Initially, the DNA sequence
is applied on the SWNT, e.g., the cytosine base attaches to the carbon atom $B_1$ and thymine
base attaches to $A_2$ and so on. When the gas is exposed to the sensor for a duration of 50 s,
then at time $\bar{t}_1$ s the gas molecule interacts with the cytosine base through vdW forces
and/or mutual polarization \cite{27}. As a result, the electron hops from the $\pi$ orbital
of one carbon atom $A_1$ to the neighboring carbon atom $B_1$ and the tunnelling of the
electron from $A_1$ to $B_1$ is an elastic process with the corresponding integral referred
to as the hopping integral $\gamma_{11}(\bar{t}_1)$. Then, the electron hops from $B_1$ to
$A_2$ with the hopping integral $\gamma_{12}(\bar{t}_1)$, and from $A_2$ to $B_2$ with
$\gamma_{22}(\bar{t}_1)=\gamma_0$, where $\gamma_0$ is the hopping integral without the gas.
Hence, the hopping integrals between other carbon atoms $\gamma_{ij}(\bar{t}_1)$ are also
$\gamma_0$, figure 2(a). The indices $i(j)$ =1,2,$\cdots p(q)$ where $p= q= M/2 $ for even
$M$ and $p= {(M-1)}/2$ and $q={(M+1)}/2$ for odd $M$. The timescale of electron transport
through the SWNT is far less than the experimental timescale of the gas flow, therefore the
electron sees a steady state at $\bar{t}_1$ s. Similarly, at $\bar{t}_2$ s another gas molecule
interacts with the thymine base attached to the carbon atom $A_2$, resulting in the hopping
integrals $\gamma_{11}(\bar{t}_2)$, $\gamma_{12}(\bar{t}_2)$, $\gamma_{22} (\bar{t}_2)$, and
other $\gamma_{ij} (\bar{t}_2)$=$\gamma_0$, figure 2(b). In this case, the electron sees a
steady state at $\bar{t}_2$ with $\gamma_{ij}(\bar{t}_2)$ different from $\gamma_{ij}(\bar{t}_1)$
when the bases interacting with the gas molecules are different. In a similar way at time
$\bar{t}_{M/2}$ for even $M$, the time when the gas molecule interacts with the last base
of the DNA sequence 2 attached to the carbon atom, the hopping integrals are $\gamma_{11}
(\bar{t}_{M/2})$, $\gamma_{12}(\bar{t}_{M/2})$, $\cdots$, $\gamma_{pq}(\bar{t}_{M/2})$,
figure 2(c). Hence, we find the hopping integrals as well as the on-site energy change
with the time $\bar{t}$ at which the gas molecules trigger the different bases of the DNA
sequence attached to the SWNT and depend on the DNA bases, gases, and geometry of attachment
of the bases and gases to the SWNT. Because of the time-dependent hopping integral and on-site
energy the relevant Green's functions for the SWNT $(G_{\rm C})$ are also time-dependent.

As the experimental timescale of the gas flow is larger than the time of electron transport
inside the SWNT ($\sim 10^{-15}-10^{-17}$s) one can express the leading transport
properties of the metal-SWNT-metal system in terms of the Green's function given by
$G(\epsilon,\bar{t})={[\epsilon-H(\bar{t})]}^{-1}$, where $\epsilon= \varepsilon \pm \rmi
\eta$ with $\rmi\eta$ an infinitesimal imaginary term and $H(\bar{t})$ is the Hamiltonian
at time $\bar{t}$, using the time-dependent NEGF formalism, an adiabatic expansion in the
slow time variable $\bar{t}=\frac{t+t{^\prime}}{2}$ and Fourier transforming with respect
to the fast variable $(t-t^{\prime})$ \cite{22,31}. The time-dependent nonequilibrium
retarded and advanced Green's functions for the SWNT can be derived as $G^{r,a}_{\rm C}
(\epsilon,\bar{t})={[{\epsilon}-H_{\rm C}(\bar{t})-{\Sigma}_{\rm S}(\epsilon,\bar{t})
-{\Sigma}_{\rm D}(\epsilon,\bar{t})]}^{-1}$, where $\Sigma_{\rm S}(\epsilon,\bar{t})$ =
${h^{\dagger}_{\rm {SC}}}g_{\rm S}(\epsilon,\bar{t}) {h_{\rm {SC}}}$ and $\Sigma_{\rm D}
(\epsilon,\bar{t})$=${h_{\rm {CD}}}g_{\rm D}(\epsilon,\bar{t}){h^{\dagger}_{\rm{CD}}}$
are the self-energy terms due to metallic contacts, with $g_{\{\rm S,\rm D\}}
(\epsilon,\bar{t})=[{\epsilon}-H_{\{\rm S,\rm D\}}(\bar{t})]^{-1}$ the Green's functions
of the contacts. This is done by expanding the Green's functions up to linear order in
the slow variable $\bar{t}$ using the adiabatic expansion \cite{22,31}:
$G^{r,a}(t-t^{\prime},\bar{t})=G^{(0)r,a}(t-t^{\prime},\bar{t})+G^{(1)r,a}
(t-t^{\prime},\bar{t})$, where $G^{(0)r,a}(t-t^{\prime},\bar{t})=G^{r,a}
(t-t^{\prime},\bar{t})|_{\bar{t}=t}$ and $G^{(1)r,a}(t-t^{\prime},\bar{t})=
(\frac{t^{\prime}-t}{2})\frac{\partial G^{r,a}}{\partial \bar{t}}
(t-t^\prime,\bar{t})|_{\bar{t}=t}$ are the zeroth and first order Green's functions.
Taking the Fourier transform with respect to the fast variable $(t-t^{\prime})$, the
Green's functions become $G^{r,a}(\varepsilon,\bar{t})=G^{(0)r,a}(\varepsilon,\bar{t})
+G^{(1)r,a}(\varepsilon,\bar{t})$. We assume that the coupling to the contacts
effectively gives rise to a finite imaginary term in the self-energies which is larger
than $\rmi\eta$ \cite{32} at all times. Therefore, we drop the term $\rmi\eta$ in the
Green's function matrix for the SWNT. The coupling functions are considered to be energy
independent $\Gamma_{\rm S/\rm D}(t,t^{\prime})=\tau(t,t^{\prime}) \times \rm exp[{i
\int^t_{t^{\prime}}dt_1 \Delta_{\rm S/\rm D}(t_1)}]$ with $\tau(t,t^{\prime})=2\pi
\sum_{\alpha \epsilon \rm S,\rm D} \rho _\alpha V_{\alpha, n}(t) V^{*}_{\alpha,n }
(t^\prime)$ \cite{25}. These functions can also be expanded using the adiabatic
expansion as $\Gamma(t-t^{\prime},\bar{t})=\Gamma^{(0)}(t-t^{\prime}, \bar{t})
+\Gamma^{(1)}(t-t^{\prime},\bar{t})$ with $\Gamma^{(0)}(t-t^{\prime},\bar{t})$=$\Gamma
(t-t^{\prime},\bar{t})|_{\bar{t}=t}=\Gamma^{(0)}(\bar{t})$=$2\pi\sum_{\alpha \epsilon
\rm S,\rm D} \rho _\alpha V_{\alpha, n}(\bar t) V^{*}_{\alpha,n }(\bar t)$ and $\Gamma^{(1)}
(\bar{t})=(t^\prime-t)\overline{\Delta} (\bar{t})= (t^{\prime}-t) \left[\frac{1}{2}
\frac{\partial \tau(t-t^{\prime},\bar{t})} {\partial{\bar{t}}} |_{\bar{t}=t}- \rmi
\tau(\bar{t})\Delta(0)\right]$ as the zeroth and first order coupling functions
\footnote{Suppressing the subscripts S and D from $\Gamma$ and $\Delta$.}. Using
these nonequilibrium Green's functions and coupling functions we can identify the
zeroth and first order contribution to the current: $I(\bar{t})=I^{(0)}(\bar{t})+
I^{(1)}(\bar{t})$.

We explicitly calculate the Green's function $(G_{\rm C})$ for the Hamiltonian
corresponding to figure 2. In the matrix form, the Hamiltonian of the system can
be divided into three blocks \cite{33} corresponding to the semiconducting SWNT
and the two metallic contacts S and D

\begin{equation}
H(\bar{t}) = \left( \begin{array}{ccc}
H_{\rm S} & h_{\rm {SC}} & 0  \\
h^\dagger_{\rm {SC}} & H_{\rm C} & h_{\rm {CD}} \\
0 & h^\dagger_{\rm {CD}} & H_{\rm D} \\
\end{array} \right),
\end{equation}

\noindent
where $H_{\{\rm S,\rm D\}}$ are the contact Hamiltonians and $H_{\rm C}$ is the
time-dependent tight-binding Hamiltonian for the SWNT with matrix elements:
$H_{A_iA_i}$=${\varepsilon_{Ai}(\bar{t})}$,$H_{B_iB_i}$=${\varepsilon_{Bi}(\bar{t})}$,
the on-site energy and $H_{A_iB_i}$=$H_{B_iA_i}$=${\gamma_{ii}(\bar{t})}$ \cite{1,34},
$H_{B_iA_{i+1}}$=$H_{A_{i+1}B_i}$=${\gamma_{i,i+1}(\bar{t})}$, the hopping integrals
and the rest of the off diagonal elements are zero. $h_{\rm {SC}}$ and $h_{\rm{CD}}$
are the coupling matrices between S and D contacts and the SWNT. Assuming the
tight-binding model and using the above Hamiltonian $H_{\rm C}(\bar t)$ the zeroth
order time-dependent retarded Green's function $G^{(0)r}_{\rm C}$ can be written
in an $M \times M$ matrix form as
\vskip 1in
\begin{eqnarray}
\fl
G^{(0) r}_{\rm C}(\varepsilon,\bar{t})=\nonumber\\
\fl
\left(
\begin{array}{cccccc}
\varepsilon-\varepsilon_{A_{1}}(\bar{t})-\Sigma_{\rm S}(\bar{t}) & {\gamma_{11}(\bar{t})} &  &  &&\\
{\gamma_{11}(\bar{t})} & \varepsilon-{\varepsilon_{B_{1}}(\bar{t})} & {\gamma_{12}(\bar{t})} & &&\\
& {\gamma_{12}(\bar{t})} & \varepsilon-{\varepsilon_{A_{2}}(\bar{t})} & \ddots&& \\
&& \ddots & \ddots &\ddots &\\
&&&\ddots&\varepsilon-\varepsilon_{X_i}(\bar{t})&\gamma_{pq}(\bar{t})\\
&&&&\gamma_{pq}(\bar{t})&\varepsilon-\varepsilon_{X_i}(\bar{t})-\Sigma_{\rm D}(\bar{t})
\end{array} \right)^{-1}.
\end{eqnarray}

\section{Results and discussion}
The zeroth order Green's functions $(G^{(0){<,r,a}})$ lead to the zeroth order current
$I^{(0)}$, which is the Landauer formula \cite{22,24,30} in terms of the slow time
variable $\bar{t}$. The Landauer formula is derived by applying the adiabatic expansion
and taking the Fourier transform of the Green's functions and the coupling functions in
the expression of the time-dependent current equation (6) of \cite{25} and considering
only the zeroth order contribution. Hence, the time-dependent Landauer formula is found
to be
\begin{equation}
I^{(0)}(\bar{t})=\frac{e}{\hbar}\int{\frac{\rmd\varepsilon}{2\pi}}~
\Tr({\Gamma^{(0)}_{\rm S}}(\bar{t}) {G^{(0){r}}_{\rm C}(\varepsilon,\bar{t})}
{\Gamma^{(0)}_{\rm D}}(\bar{t}) {G^{(0){a}}_{\rm C}(\varepsilon,\bar{t})})
[f_{\rm S}(\varepsilon)-f_{\rm D}(\varepsilon)].
\end{equation}
This is a significant result of this paper.

The transmission function of the system is identified as $T(\varepsilon,\bar{t})=\Tr({\Gamma^{(0)}
_{\rm S}}(\bar{t}){G_{\rm C}^{(0){r}}(\varepsilon,\bar{t})}{\Gamma^{(0)}_{\rm D}}(\bar{t}){G_{\rm C
}^{(0){a}}(\varepsilon,\bar{t})})$, where $G_{\rm C}^{(0){r,a}}(\varepsilon, \bar{t})$ are
the zeroth order retarded and advanced Green's functions of the SWNT at $\bar{t}$ and
$f_{\{\rm S,\rm D\}}(\varepsilon)$ are the Fermi distribution functions in the source
and drain contacts. The coupling functions are related to the self-energies by the
relationship $\Gamma_{\rm \{S,D\}}(\bar{t})=\rm i[\Sigma_{\{S,D\}}(\bar{t})-\Sigma^
\dagger_{\{S,D\}}(\bar{t})]$, where we have considered self-energies to be energy
independent. $\rm \Sigma_{\{S,D\}}(\bar{t})$ contains both real and imaginary parts.

To calculate the transmission function, we have assumed that only the first element of
the $\Gamma^{(0)}_{\rm S}(\bar{t})(\Sigma_{\rm S}(\bar{t}))$ matrix and the last element
of the $\Gamma^{(0)}_{\rm D}(\bar{t}) (\Sigma_{\rm D}(\bar{t}))$ matrix are present. The
rest of the elements of matrices are considered to be zero. Then, the transmission function
depends only on one off diagonal element of $G_{\rm C}$: $T(\varepsilon,\bar{t})={\Gamma^{(0)}
_{\rm S,11}(\bar{t})}{G^{(0)}_{\rm C1M}(\varepsilon,\bar{t})}{\Gamma^{(0)}_{\rm D,MM}(\bar{t})}
{G^{(0){*}}_{\rm C1M}(\varepsilon,\bar{t})}$.

To get an explicit expression for the current, we consider linear response \cite{32,33}.
In linear response, equation (4) becomes $I^{(0)}(\bar{t})=\frac{e}{\hbar}\int{\frac{\rmd
{\varepsilon}}{2\pi}}T({\varepsilon},\bar{t})\delta[f(\varepsilon-\mu_{\rm S})-f(\varepsilon-
\mu_{\rm D})]$, where $\mu_{\{\rm S,\rm D\}}$ are the chemical potentials associated with
the source and drain contacts. This equation then can be written as $I^{(0)}(\bar{t})=\frac
{e}{\hbar}T({\varepsilon_f},\bar{t})[\mu_{\rm S}-\mu_{\rm D}]$ as $\delta[f(\varepsilon-\mu_
{\rm S})-f(\varepsilon-\mu_{\rm D})]=(\mu_{\rm S}-\mu_{\rm D})(-\frac{\partial f}{\partial
\varepsilon})$ and $(-\frac{\partial f}{\partial\varepsilon})=\delta(\varepsilon_f-\varepsilon)$
\cite{32} where $\varepsilon_f$ is the Fermi energy.

In the experiment \cite{12}, there are two DNA sequences: sequence 1 with 21 bases and
sequence 2 with 24 bases. Hence, we have two matrices, one is a $23 \times 23$ matrix
and the other is a $26 \times 26$ matrix \footnote{Since the first and the last carbon
atoms are connected to the source and drain contacts.}. For the above Hamiltonian matrix
and using $\Gamma^{(0)}_{\rm S,11}(\bar{t})=-2 Im(\Sigma_{\rm S,11}(\bar{t}))=-2 Im\Sigma_
{\rm S}(\bar{t})$ and $\Gamma^{(0)}_{\rm D,MM}(\bar{t})= -2 Im(\Sigma_{\rm D,MM}(\bar{t}))=
-2 Im\Sigma_{\rm D}(\bar{t})$, the general expression for the transmission function of a
SWNT decorated with DNA sequence 1 (2) and gas is given by\footnote{Note: equation (5) for
$\gamma_{ij}=\gamma_0$ and $\varepsilon_{X_i}=\varepsilon_f=0$ reduces to equations (10)
and (11) in \cite{33}.}
\begin{equation}
T(\varepsilon_f,\bar{t})=\frac{4~Im{\Sigma_{\rm S}(\bar{t})}~Im{\Sigma_{\rm D}
(\bar{t})}~{\gamma^{2}_{11}(\bar{t})}{\gamma^{2}_{12}(\bar{t})}\cdots{\gamma^{2}_{{pq}}}
(\bar{t})}{|{G_{\rm C}}^{(0)r}(\varepsilon_f,\bar{t})|^2_{M \times M}}.
\end{equation}
Equation (5) gives an explicit formula for the transmission function in terms of
$\gamma_{ij}$ and $\varepsilon_{X_i}$ indicating the dependence of the transmission
function and hence the current on the hopping integrals and the on-site energies, which
are functions of time $\bar{t}$.

We also derive the first order contribution to the time-dependent current using
equation (6) of \cite{25}:
\vskip -.2in
\begin{eqnarray}
I^{(1)}_{\rm S/\rm D}(\bar{t})&=& - \frac{e}{\hbar}Im \Tr \Bigg [
\Gamma^{(0)}(\bar{t}) \bigg \{\sum_{\rm S,\rm D}\int
\frac{\rmd\varepsilon}{2\pi} f(\varepsilon) \bigg \{ \bigg(
\frac{\partial} {\partial \varepsilon}
\{G^{(0)r} G^{(0)a}\}\bigg)
\overline{\Delta}(\bar{t})\nonumber\\
&+& \rmi\bigg(G^{(0)r}G^{(1)a} + G^{(1)r}G^{(0)a}\bigg)
\Gamma^{(0)}(\bar{t}) \bigg \} \bigg \}
\Bigg ]\nonumber\\
&-& \frac{e}{\hbar} Im Tr \Bigg[ \int \frac{\rmd\varepsilon}{2\pi}
f(\varepsilon) \bigg\{\Gamma^{(0)} (\bar{t})G^{(1)r}-\rmi
\overline{\Delta}(\bar{t})\frac{\partial G^{(0)r}}{\partial
\varepsilon} \bigg \} \Bigg ],
\end{eqnarray}
\noindent
where we have used $\frac{1}{\rmi}\frac{\partial}{\partial\varepsilon}G^{(0)r,a}
(\varepsilon,\bar{t})=\int^\infty_{{-}\infty} \rmd(t-t^{\prime})e^{\rmi\varepsilon
(t-t^\prime)}(t-t^\prime)G^{(0)r,a}(t-t^\prime,\bar{t})$ and the dependence of
$G^{(0)r,a}$ and $G^{(1)r,a}$ on $\varepsilon$ and $\bar{t}$ has been suppressed.
An expression for $G^{(1)r,a}(\varepsilon, \bar{t})$ can be explicitly calculated
from $G^{r,a}(t-t^\prime,\bar{t})$ and equation (3) using the formula:
$G^{(1)r,a}(\varepsilon,\bar{t})={-}(\frac{t^\prime-t}{2})G^{(0)r,a}
(\varepsilon,\bar{t})\frac{\partial}{\partial\bar{t}}G^{-1 r,a}
(t-t^\prime,\bar{t})|_{\bar{t}=t}G^{(0)r,a}(\varepsilon,\bar{t})$. This result is a
contribution in addition to the Landauer formula for the current, equation (4), and has
been presented for the first time for such gas sensors.

The interaction of different gas molecules with DNA-decorated SWNT causes a redistribution
of the charge in the system, leading to a partial charge transfer from the Gas-DNA-base
complex to the SWNT. This deforms the SWNT and changes the nearest-neighbour carbon-carbon
distance $\rm a_{cc}$, thus affecting the hopping integral, i.e., hopping of electrons
between the adjacent carbon atoms and the on-site energy and therefore changes the sensor
response. To analyse the sensor response in terms of the hopping integral and on-site
energy let us consider the model parameters $\gamma_{ij}$ and $\varepsilon_{X_i}$ to
have the form $\gamma_{ij}(\rm \Delta a_{cc}(\bar{t}))=\gamma_0 \rm exp(-\Delta a_{cc}
(\bar{t})/{a_0})$ and $\varepsilon_{X_i}(\rm \Delta a_{cc}(\bar{t}))=\varepsilon_0 \rm
exp(-\Delta a_{cc}(\bar{t})/{a_0})$ \cite{35}. Here $\gamma_0$ and $\varepsilon_0$ are
the hopping integral and on-site energy without the gas, with $\rm a_0=0.33$ \AA. $\gamma_{ij}
(\rm \Delta a_{cc}(\bar{t}))$ and $\varepsilon_{X_i}(\rm \Delta a_{cc}(\bar{t}))$ are the
modified parameters when the nearest-neighbour carbon-carbon distance $\rm a_{cc}$ changes
by $\rm \Delta a_{cc}$ as a function of time $\bar{t}$ due to the interaction at time
$\bar t$ of gas molecules with the DNA-decorated SWNT.

The experimental observations \cite{12} show that the current decreases when the device
is exposed to different gases, apart from PA, and it increases when exposed to air. To demonstrate
that the theory and experiment are in agreement we explicitly calculate the values for
$\Delta I^{(0)}(\bar{t})/{I_0}$ for a 7 $\times$ 7 matrix (for simplicity of the calculation)
using equations for the current (4), the transmission function (5) and the form for $\gamma_{ij}$
in terms of $\rm \Delta a_{cc}(\bar t)$ (given in the above paragraph) for methanol with
DNA sequence 2 and air. Here, $I_0$ is the current without the odour. To reproduce the sensor
response we fix the parameters $\rm \Delta a_{cc}(\bar t)$ and $\varepsilon_f(\bar t)$. Table
1 shows the values of these parameters at each time $\bar t$ considered in the calculation,
where we find $\rm \Delta a_{cc}(\bar t)$ and $\varepsilon_f(\bar t)$ are sensitive to the
bases of the DNA sequence. We observe that when the value of $\rm \Delta a_{cc}(\bar t)$
decreases (negative value) and $\varepsilon_f(\bar t)$ increases the corresponding sensor
current decreases. The presence of gas molecules and DNA causes a charge transfer from each
Gas-DNA-base complex to the SWNT, which decreases $\rm a_{cc}$ from its pristine values (when
there is no gas) and increases $\varepsilon_f(\bar t)$ at each time. This enhances the
hopping integral $(\gamma_{ij}=\gamma_0 \rm exp(\Delta a_{cc}(\bar{t})/{a_0}))$ and hence
the transmission function, equation (5). As a result, the electron current increases reducing
the hole current of the p-type SWNT. An increase in $\varepsilon_f(\bar t)$ indicates that
the Fermi level shifts away from the valence band (here we have assumed only the contribution
of $\gamma_{ij}$ and neglected $\varepsilon_{X_i}$in (5)). Figure 3 is the plot between the
sensor response, calculated using the parameters given in table 1, and the time for methanol.
The sensor response at times $\bar t_2, \bar t_3$ and $\bar t_5$ indicates that the Gas-thymine-base
complex transfers charge to the SWNT, causing a decrease in $\rm a_{cc}$ and an increase in
$\varepsilon_f$, resulting in a decrease in the hole current. While the response at times
$\bar t_1$ and $\bar t_4$ indicates that the Gas-cytosine-base complex transfers less charge
to the SWNT, causing less decrease in $\rm a_{cc}$ and less increase in $\varepsilon_f$ (compared
to Gas-thymine-base complex), resulting in an increase in the hole current. This sensor response
is found to be consistent with the experimental result and shows its sensitivity to the DNA
bases.

On the other hand, table 2 gives the values of the parameters $\rm \Delta a_{cc}(\bar t)$
and $\varepsilon_f(\bar t)$ when the DNA-SWNT sensor is exposed to air. When the gas
molecules are replaced by air molecules the charge transfer takes place from SWNT
to the Air-DNA-base complex, causing an increase in $\rm a_{cc}(\bar t)$ (positive value)
from its modified value due to gas and a decrease in $\varepsilon_f(\bar t)$. The zero
value of $\rm \Delta a_{cc}(\bar t)$ indicates that $\rm a_{cc}$ increases and acquires
its pristine value (figure 4). This results in lowering the hopping integral
$(\gamma_{ij}=\gamma_0\rm exp(-\Delta a_{cc}(\bar{t})/{a_0}))$ and hence the transmission
function, leading to a decrease in the electron current and therefore an increase in
the hole current. A decrease in $\varepsilon_f(\bar t)$ shows that the Fermi level moves
towards the valence band. Figure 4 shows the changes in $\rm a_{cc}(\bar t)$ when air
molecules replace the gas molecules. Figure 5 gives the sensor response to air, where the
charge transfer from SWNT to the Air-thymine-base complex is larger than the charge
transfer from SWNT to the Air-cytosine-base complex. This indicates the sensitivity of
the sensor response to the DNA bases. Figures 3 and 5 show a good match
between the experimental and theoretical results. Hence, the formula reproduces the
current characteristics of the experiment \cite{12}.

Using $I(\bar{t})$ the net charge transfer per second can be calculated as $\Delta
Q(\bar{t}_i)=I({\bar{t}_i})-I(\bar{t}_{i-1})$. Using the expressions for $G^{(0)}$
and $G^{(1)}$, we find that $I^{(1)}(\bar{t})$ changes with $\gamma_{ij}(\bar{t})$,
$\varepsilon_{X_{i}}(\bar{t})$, $\Sigma_{\rm S,\rm D}(\bar{t})$, $\overline{\Delta}
(\bar {t})$ and $\frac{\partial\gamma_{ij}(\bar{t})}{\partial\bar{t}}$, $\frac{\partial
\varepsilon_{X_{i}}(\bar{t})}{\partial\bar{t}}$, $\frac{\partial\Sigma_{\rm S,\rm D}
(\bar{t})}{\partial\bar{t}}$ \footnote{For $\overline{\Delta}(\bar{t})=0$, equation (6)
for $I^{(1)}(\bar{t})$ depends on $G^{(1)}$ and hence is small compared to $I^{(0)}
(\bar{t})$.} while $I^{(0)}(\bar{t})$ changes with $\gamma_{ij}(\bar{t})$, $\varepsilon
_{X_{i}}(\bar{t})$ and $\Sigma_{\rm S,\rm D}(\bar{t})$. These are some predictions of
the model for the experiment. The results for the time-dependent current can be used
to calculate the two-time current correlation function and compared with the plots
calculated from the experiment \cite{12}, figure 1 and will be reported elsewhere.

\section{Conclusions}
In conclusion, the two-time current correlation function for the experimental data
\cite{12} has been calculated and for this system analytical calculations of quantum
electronic transport have been presented by setting up a tight-binding model and
applying the time-dependent NEGF formalism. The Green's functions and coupling functions
have been expanded using the adiabatic expansion in the slow variable and the Fourier
transform has been taken with respect to the fast variable. With the help of the Green's
functions and coupling functions, the zeroth and first order contributions to the current
have been investigated. We explicitly calculate the sensor response by considering a
form for the hopping integral in terms of $\rm \Delta a_{cc}(\bar t)$. The sensor
response is found to be sensitive to the DNA bases.

Equations (1), (4)-(6) carry the principal results of this paper. The correlation function
equation (1) is a measure of the sensitivity and selectivity of the DNA-decorated SWNT
sensors and suggest that these gas flow sensors may also be used as DNA sequence detectors,
where the pattern of correlation functions may be used as a benchmark for the particular
chemical signal encoded in a DNA sequence. Equation (4) presents the zeroth order time-dependent
current, which is the Landauer formula that depends on the slow time variable $\bar{t}$. The
dependence of the transmission function and hence the current, on $\gamma_{ij}(\bar{t})$ and
$\varepsilon_{X_i} (\bar{t})$, has been shown by an explicit formula, equation (5). Equation
(6) shows the first order contribution to the current for the experiment \cite{12}, which is
proportional to the zeroth and first order Green's functions. The formula for the time-dependent
current is then used to compare the theoretical results with the experiment. An expression
for the net charge transfer per second is obtained using the current.

The numerical and analytical approaches used in this work can be applied to a broad range
of systems where gas flows over nano-structures doped with different chemical and biological
molecules. This will provide a method for the study of time-dependent electronic transport
in low dimensional disordered systems with gas flow. We believe that the analyses done in
this manuscript are also applicable to DNA-decorated graphene sensors \cite{36} and give
predictions to strengthen future experiments.

\section*{Acknowledgments}
We would like to thank Professors A T Charlie Johnson, M Muller, A Silva, V Falko, J
Robinson, and Deepak Kumar for helpful discussions and suggestions. We also thank the
Council of Scientific and Industrial Research (CSIR), the University Grant Commission (UGC),
and the University Faculty R $\&$ D Research Programme for financial support.

\newpage
\begin{figure}
\center
\includegraphics[width=8.5cm]{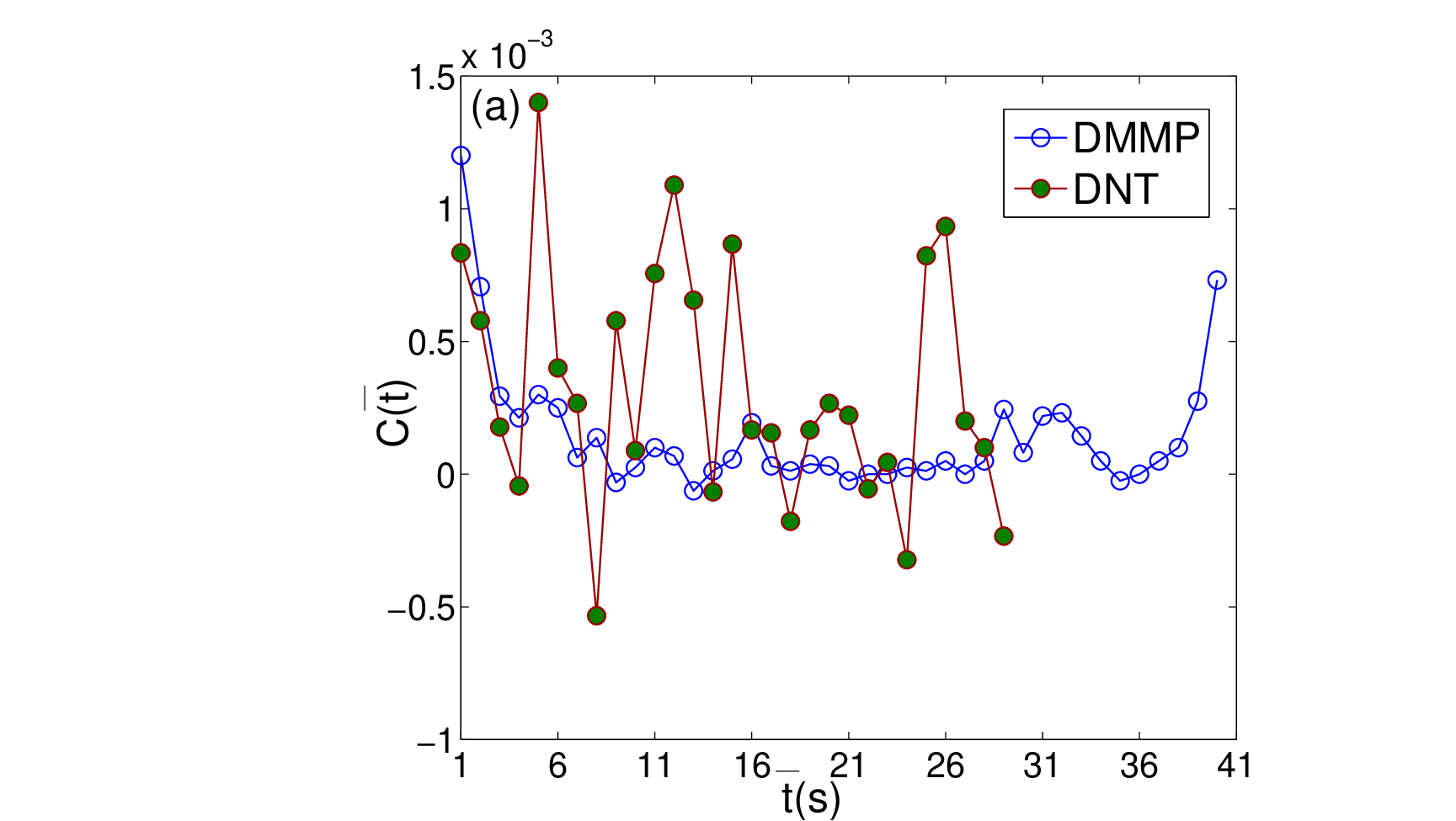}\includegraphics[width=8.5cm]{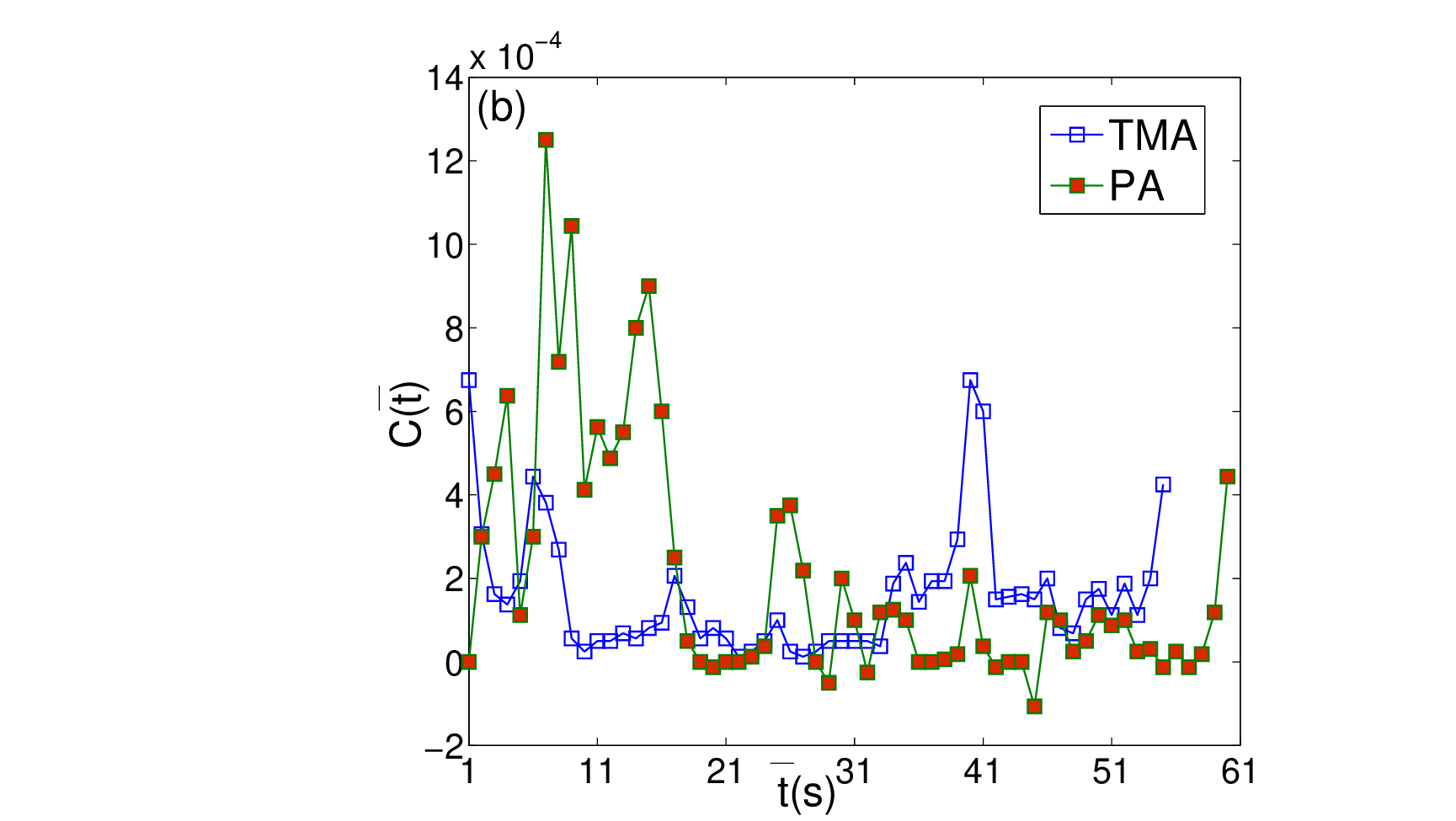}
\includegraphics[width=8.5cm]{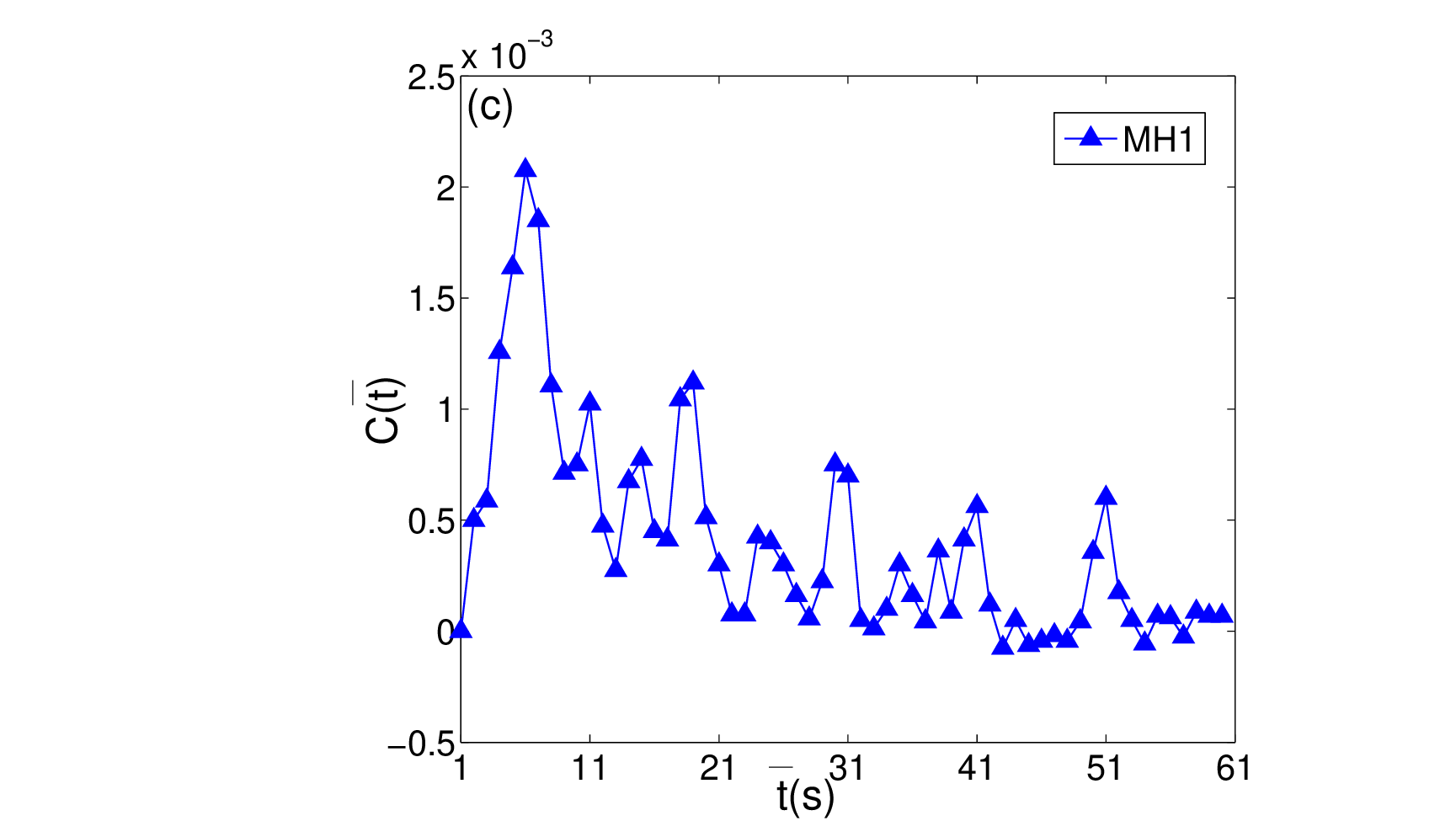}\includegraphics[width=8.5cm]{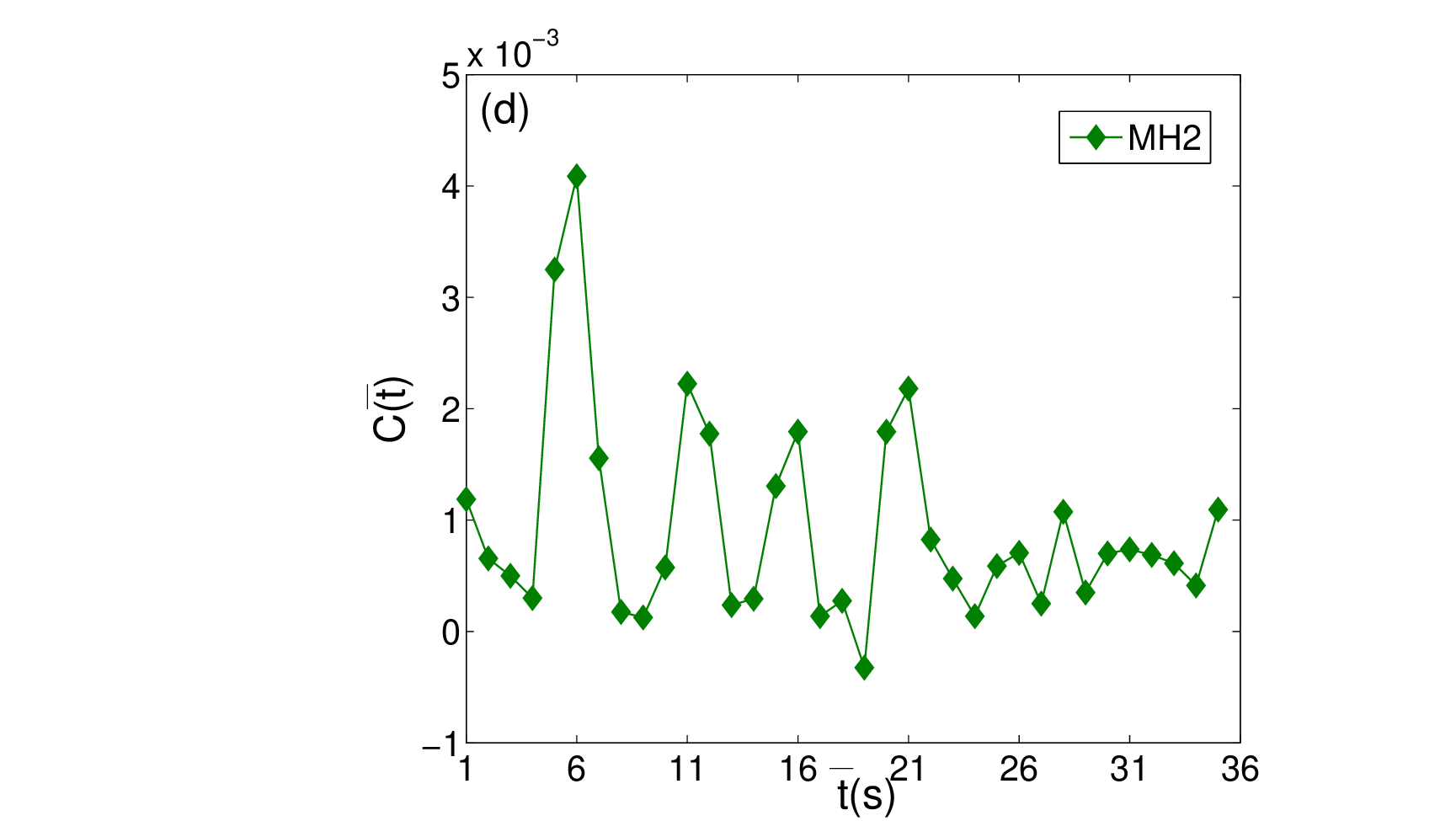}
\caption{\small{The two-time current correlation function $C(\bar{t})$ versus $\bar{t}$(s)
for (a) DMMP + sequence 2 and DNT + sequence 1 (b) TMA + sequence 2 and PA + sequence 1 (c)
methanol + sequence 1 and (d) methanol + sequence 2.}}
\end{figure}

\newpage
\begin{figure}
\center
\includegraphics[width=8cm]{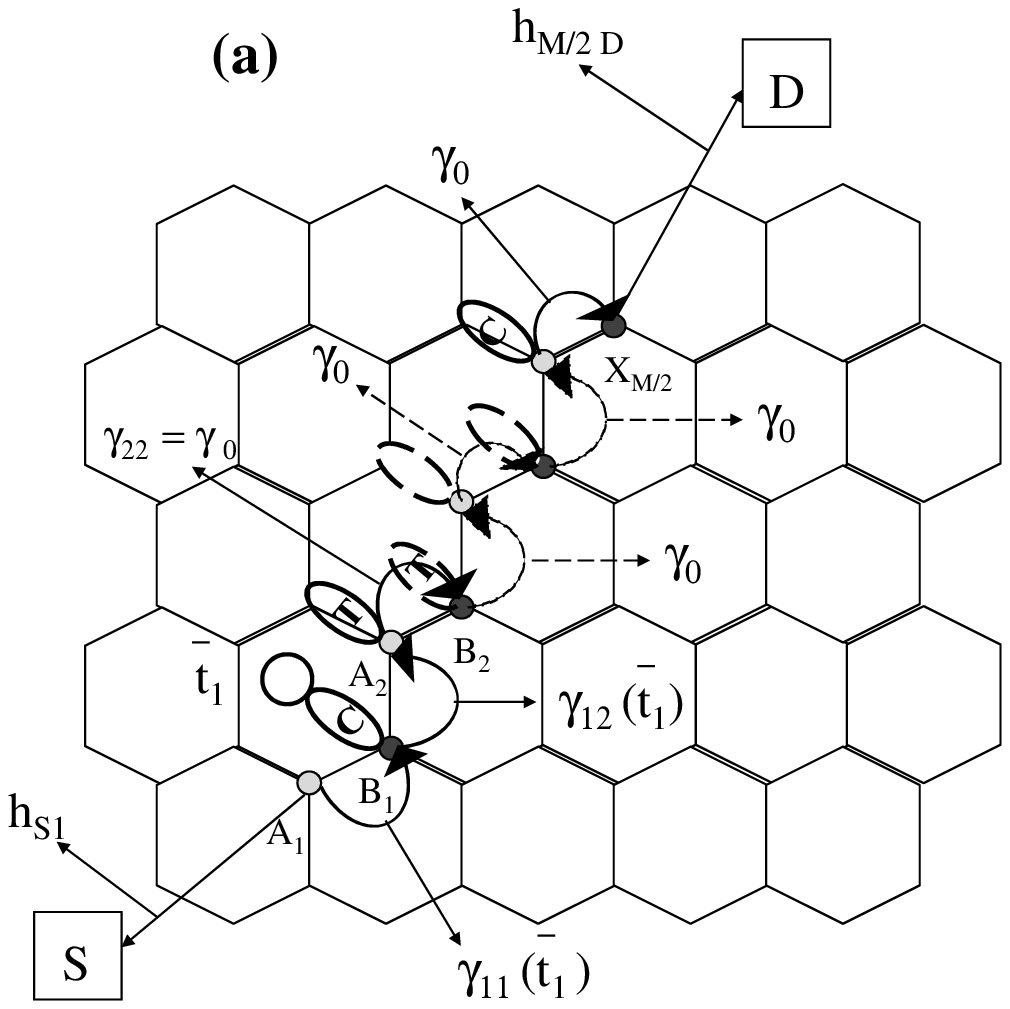}\includegraphics[width=8cm]{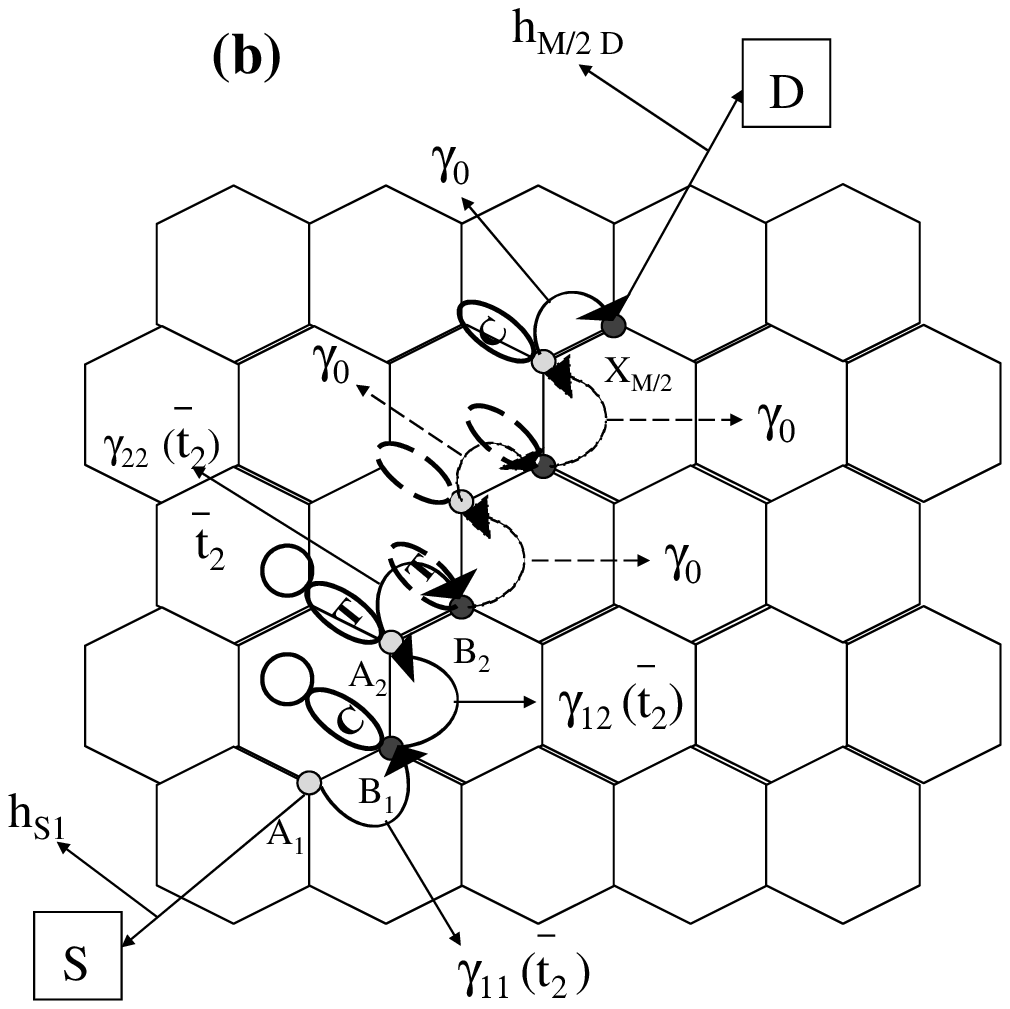}
\includegraphics[width=8cm]{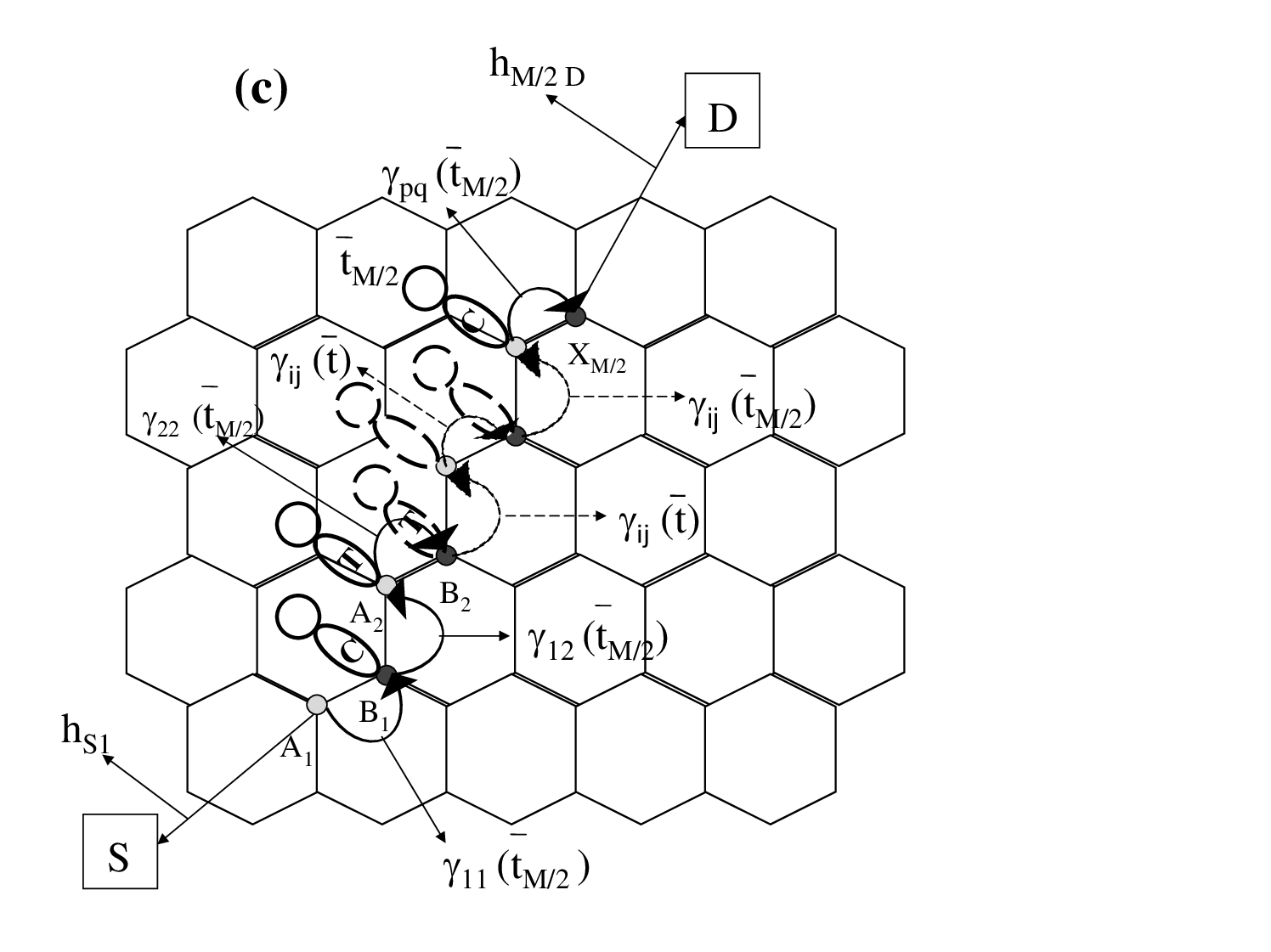}
\vspace{-0.5cm}
\caption{\small{Portion of an unwrapped SWNT. ${\gamma_{ij}(\bar{t})}$ are the hopping
integrals indicating hopping between different carbon atoms $A_i$ and $B_i$ at time
$\bar{t}$. Here, $\bar{t} = \bar{t}_1, \bar{t}_2, \cdots, \bar{t}_{M/2}$. $\gamma_0$
represents the hopping integral in the absence of the gas. $h_{\rm S1}$ represents
the coupling between the source contact and the first carbon atom $A_1$ while $h_{M/2
\rm ~D}$ is the coupling between the last carbon atom $X_{M/2}$ and the drain contact.
The ovals represent the bases cytosine, thymine etc. of DNA sequence 2 while the
circles indicate the gas molecules.}}
\end{figure}

\begin{table}
\caption{\label{tabone}Values for different parameters used in the calculation of
the current with $\gamma_0$=2.5eV for methanol with DNA sequence 2. The units of
$\varepsilon_f$ and $\rm \Delta a_{cc}$ are eV and \AA ~respectively.}
\begin{indented}
\lineup
\item[]\begin{tabular}{@{}*{6}{l}}
\br
$\0\0\0\bar{t}(s)$&\m$\bar{t_1}$&\m$\bar{t_2}$&\m$\bar{t_3}$&\m$\bar{t_4}$&\m$\0\bar{t_5}$\cr
\mr
\0\0$\Delta I^{(0)}(\bar{t})/{I_0}$&$0.0151$&$-0.0727$&$-0.08197$&$-0.0728$&$-0.1119$\cr
\0\0$\varepsilon_f(\bar{t})$&\m1.68&\m3.29&\m3.35&\m1.83&\m2.96\cr
\0\0$\rm \Delta a_{A_1B_1}(\bar{t})$&$-0.004$&$-0.004$&$-0.004$&$-0.004$&$-0.004$\cr
\0\0$\rm \Delta a_{B_1A_2}(\bar{t})$&$-0.006$&$-0.012$&$-0.012$&$-0.012$&$-0.012$\cr
\0\0$\rm \Delta a_{A_2B_2}(\bar{t})$&\m--- &$-0.006$&$-0.012$&$-0.012$&$-0.012$\cr
\0\0$\rm \Delta a_{B_2A_3}(\bar{t})$&\m--- &\m--- &$-0.004$&$-0.008$&$-0.008$\cr
\0\0$\rm \Delta a_{A_3B_3}(\bar{t})$&\m--- &\m--- &\m--- &$-0.006$&$-0.012$\cr
\0\0$\rm \Delta a_{B_3A_4}(\bar{t})$&\m--- &\m--- &\m--- &\m--- &$-0.006$\cr
\br
\end{tabular}
\end{indented}
\end{table}

\begin{figure}[b]
\center
\includegraphics[width=8cm]{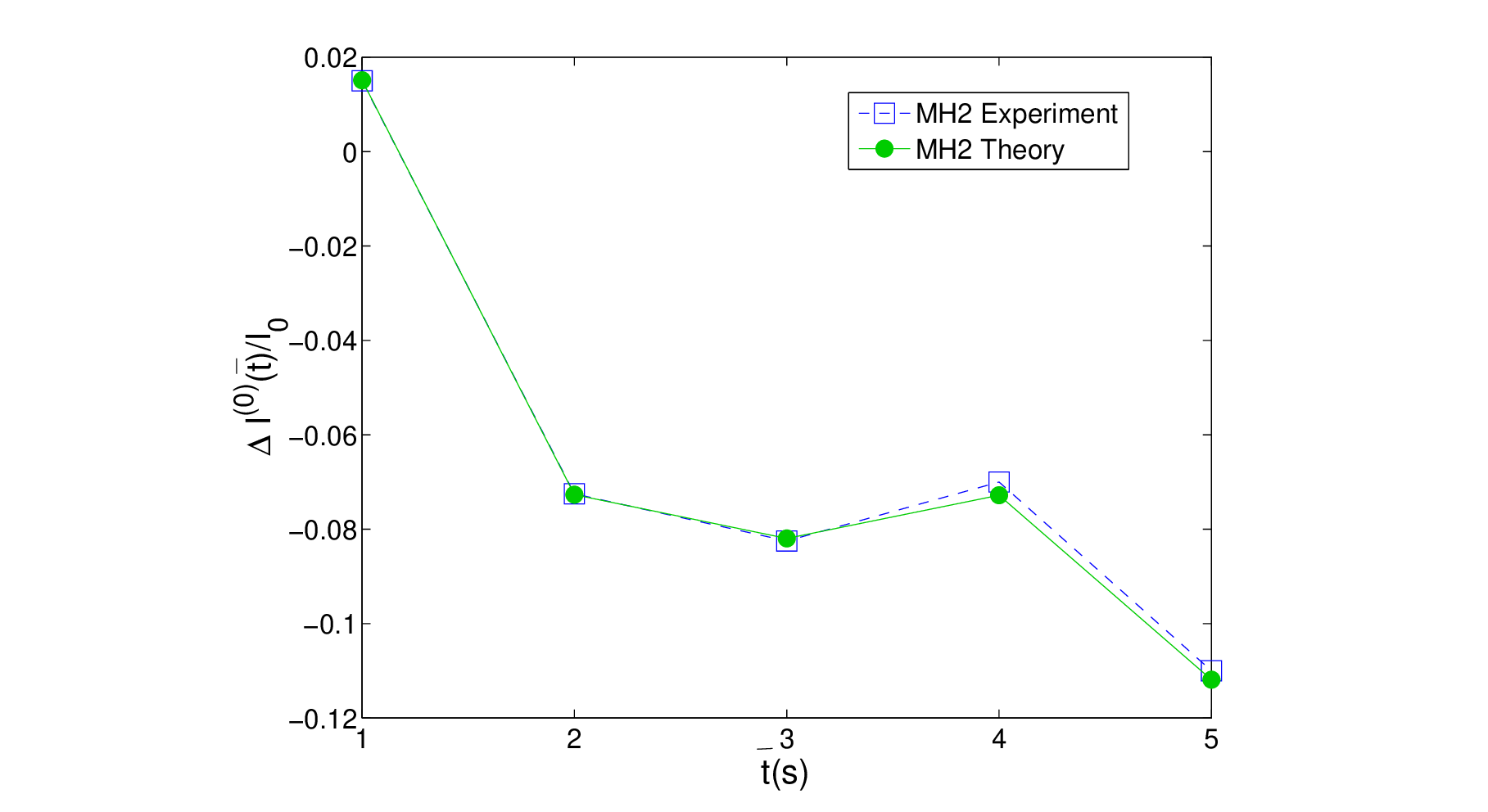}
\vspace{-0.5cm}
\caption{\small{The sensor response $\Delta I^{(0)}(\bar{t})/{I_0}$
versus time $\bar{t}$(s) for methanol with DNA sequence 2. This plot shows agreement
between the theory and experiment. We have found similar results for methanol with
the other DNA sequence 1. Experimental data is used from \cite{12} with permission
from American Chemical Society.}}
\end{figure}

\newpage
\begin{figure}
\center
\includegraphics[width=9cm]{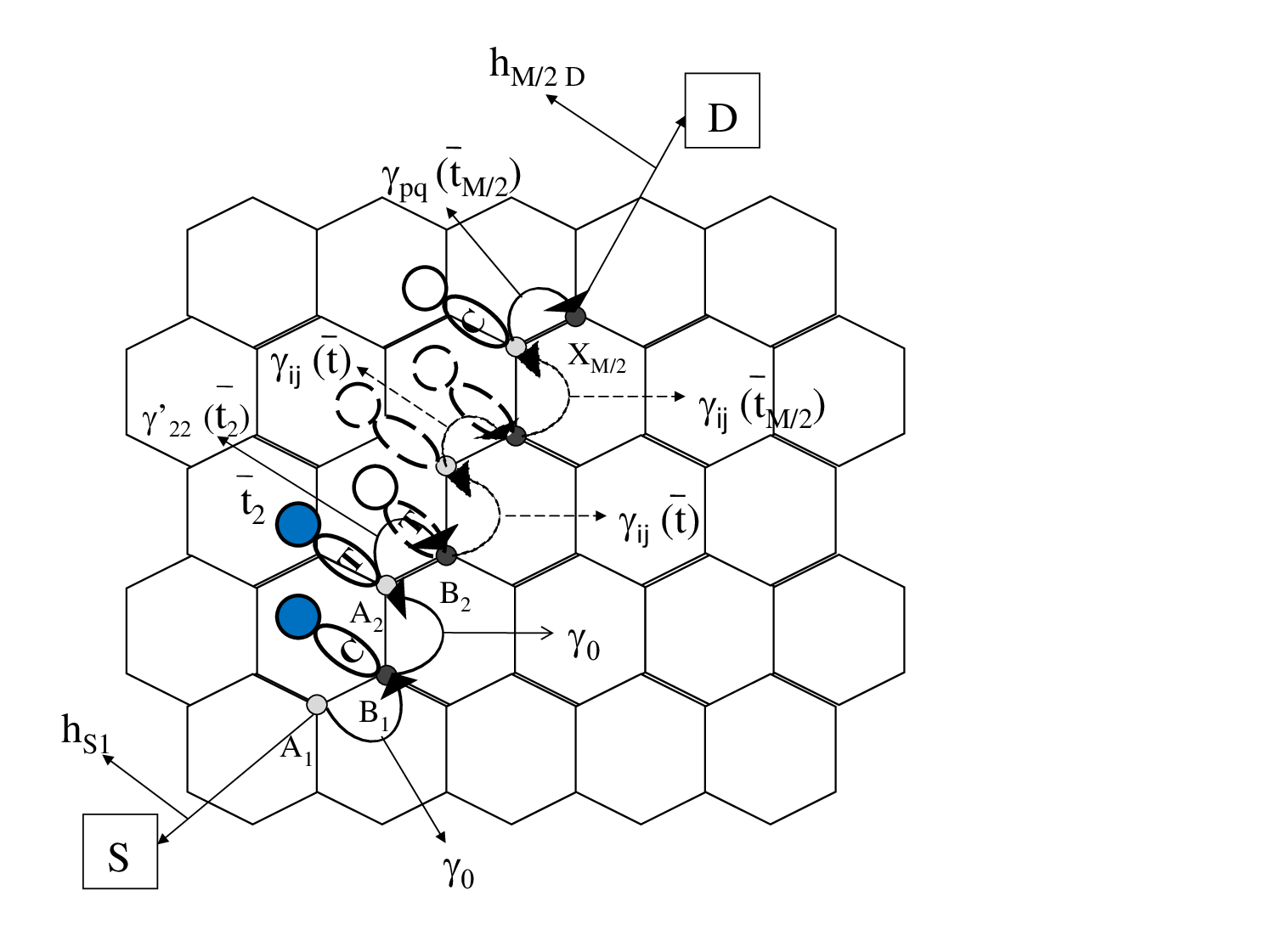}
\caption{\small{When the device is exposed to air $\rm a_{cc}$ changes. $\rm \Delta a_{A_1B_1}
(\bar t_2)$ and $\rm \Delta a_{B_1A_2}(\bar t_2)$ become zero, leading to ${\gamma_{11}(\bar{t_2})
}$=${\gamma_{12}(\bar{t_2})}$=$\gamma_0$, and $\rm\Delta a_{A_2B_2}(\bar t_2)$ increases, resulting
in ${\gamma_{22}^\prime(\bar{t_2})}$, which is different from ${\gamma_{22}(\bar{t_2})}$ (figure 2(b)).
The solid circles represent the air molecules.}}
\end{figure}

\begin{table}
\caption{\label{tabone}Values for different parameters used in the calculation of the
current with $\gamma_0$=2.5eV for air.}
\begin{indented}
\lineup
\item[]\begin{tabular}{@{}*{6}{l}}
\br
$\0\0\0\bar{t}(s)$&\m$\bar{t_1}$&\m$\bar{t_2}$&\m$\bar{t_3}$&\m$\bar{t_4}$&\m$\0\bar{t_5}$\cr
\mr
\0\0$\Delta I^{(0)}(\bar{t})/{I_0}$&$-0.2008$&$-0.1679$&$-0.1327$&$-0.1157$&$-0.1150$\cr
\0\0$\varepsilon_f(\bar{t})$&\m3.55&\m3.34&\m3.22&\m1.48&\m2.19\cr
\0\0$\rm \Delta a_{A_1B_1}(\bar{t})$&\m0&\m0&\m0&\m0&\m0\cr
\0\0$\rm \Delta a_{B_1A_2}(\bar{t})$&\m0.020&\m0&\m0&\m0&\m0\cr
\0\0$\rm \Delta a_{A_2B_2}(\bar{t})$&\m-0.012&\m0.020&\m0&\m0&\m0\cr
\0\0$\rm \Delta a_{B_2A_3}(\bar{t})$&\m-0.008&\m-0.008&\m0.014&\m0&\m0\cr
\0\0$\rm \Delta a_{A_3B_3}(\bar{t})$&\m-0.012&\m-0.012&\m-0.012&\m0.020&\m0\cr
\0\0$\rm \Delta a_{B_3A_4}(\bar{t})$&\m-0.006&\m-0.006&\m-0.006&\m-0.006&\m0.014\cr
\br
\end{tabular}
\end{indented}
\end{table}

\begin{figure}[h]
\center
\includegraphics[width=9cm]{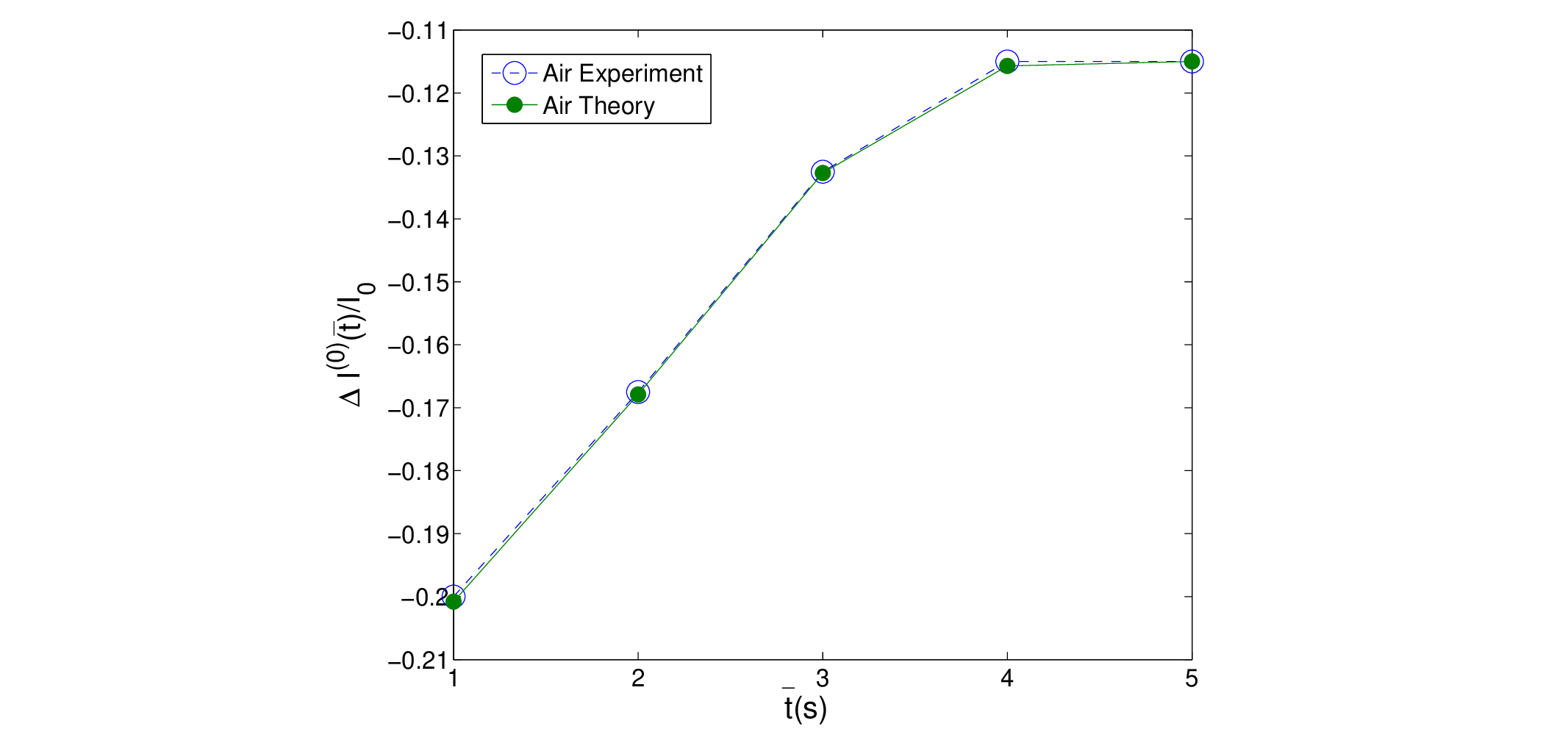}
\caption{\small{The sensor response $\Delta I^{(0)}(\bar{t})/{I_0}$ versus time $\bar{t}$(s)
for air. This plot again shows agreement between theory and experiment.}}
\end{figure}

\newpage

\section*{References}

\begin{thebibliography}{00}

\bibitem{1}
Saito R,  Dresselhaus G and Dresselhaus M S 1998 {\it Physical
Properties of Carbon Nanotubes} (London: Imperical College Press)

\bibitem{2}
Kong J, Franklin N R, Zhou C, Chapline M G, Peng S, Cho K and Dai H
2000 \emph{Science} \textbf{287} 622

\bibitem{3}
Collins P G,  Bradley K, Ishigami M and Zettl A
2000 \emph{Science} \textbf{287} 1801

\bibitem{4}
Li J, Lu Y, Ye Q, Cinke M, Han J and Meyyappan M
2003 \emph{Nano Lett.} \textbf{3} 929

\bibitem{5}
Snow E, Perkins F K
2005 \emph{Nano Lett.} \textbf{5} 2414

\bibitem{6}
Latil S , Roche S and Charlier J C
2005 \emph{Nano Lett.} \textbf{5} 2216

\bibitem{7}
Snow E S, Perkins F K, Houser E J, Badescu S C and Reinecke T L
2005 \emph{Science} \textbf{307} 1942

\bibitem{8}
Shim M, Javey A,  Kam N W S and Dai H
2001 \emph{J. Am. Chem. Soc.} \textbf{123} 11512

\bibitem{9}
Qi P, Vermesh O, Grecu M, Javey A, Wang Q, Dai H, Peng S and Cho K J
2003\emph{ Nano Lett.} \textbf{3} 347

\bibitem{10}
Peng S and Cho K
2003 \emph{Nano Lett.} \textbf{3} 513

\bibitem{11}
Latil S, Roche S, Mayou D and Charlier J C
2004 \emph{Phys. Rev. Lett.} \textbf{92} 256805

\bibitem{12}
Staii C, Johnson Jr A T, Chen M and Gelperin A 2005 \emph{Nano Lett.} \textbf{5} 1774

\bibitem{13}
Zheng M, Jagota A, Semke E D, Diner B A, Mclean R S, Lustig S R, Richardson R E and Tassi N G
2003 \emph{Nat. Mater.} \textbf{2} 338

\bibitem{14}
Zheng M et al 2003 \emph{Science} \textbf{302} 1545

\bibitem{15}
Jeng E S, Moll A E, Roy A C, Gastala J B and Strano M S,
2006 \emph{Nano Lett.} \textbf{6} 371

\bibitem{16}
Tu X, Manohar S, Jagota A and Zheng M
2009 \emph{Nature} \textbf{460} 250

\bibitem{17}
Enyashin A N, Gemming S and Seifert G
2007 \emph{Nanotechnology} \textbf{18} 245702

\bibitem{18}
Meng S, Maragakis P, Papaloukas C and Kaxiras E 2007 \emph{Nano Lett.} \textbf{7} 45

\bibitem{19}
Gowtham S, Scheicher R H, Pandey R, Karna S P and Ahuja R
2008 \emph{Nanotechnology} \textbf{19} 125701

\bibitem{20}
Johnson R R, Johnson A T C and Klein M L
2010 \emph{Small} \textbf{6} 31

\bibitem{21}
Gao H and Kong Y 2004 \emph{Annu. Rev. Mater. Res.} \textbf{34} 123

\bibitem{22}
Haug H J W and Jauho A P  2008 {\it Quantum Kinetics in Transport and
Optics of Semiconductors} (Berlin Heidelberg New York: Springer)

\bibitem{23}
Nardelli M B 1999 \emph{\PR B}  \textbf{60} 7828

\bibitem{24}
Meir Y and Wingreen N S 1992 \emph{\PRL} \textbf{68} 2512

\bibitem{25}
Wingreen N S, Jauho A P and Meir Y 1993 \emph{\PR B} \textbf{48} 8487

\bibitem{26}
Kienle D and Leonard F 2009 \emph{\PRL} \textbf{103} 026601

\bibitem{27}
Poonam P and Deo N 2008 \emph{Sens. Actuators B Chem.} \textbf{135} 327

\bibitem{28}
Imry Y 1997 {\it Introduction to Mesoscopic Physics} (Oxford: Oxford University
Press)

\bibitem{29}
Lee P A 1986 \emph{Physica} \textbf{140A} 169

Lee P A and Ramakrishnan T V 1985 \emph{Rev. Mod. Phys.} \textbf{57} 287

Lee P A and Stone A D 1985 \emph{\PRL} \textbf{55} 1622

Al'tshuler B L, Lee P A and Webb R A {\it Mesoscopic Phenomena in Solids} 1991 (New York: North-Holland)

\bibitem{30}
Datta S 2007 {\it Quantum Transport: Atom to Transistors}
(Cambridge: Cambridge University Press)

\bibitem{31}
Hernandez A R, Pinheiro F A, Lewenkopf C H and Mucciolo E
R 2009 \emph{\PR B} \textbf{80} 115311

\bibitem{32}
Datta S 1995  {\it Electronic Transport in Mesoscopic Systems}
(Cambridge: Cambridge University Press)

\bibitem{33}
Chen Y R, Zhang L and Hybertsen M S 2007 \emph{\PR B} \textbf{76} 115408

\bibitem{34}
Falko V 2006 {\it Quantum transport of chiral electrons in graphene I}
(Lecture note in `College on Physics of Nano-Devices') The Abdus
Salam International Centre for Theoretical Physics (Trieste: Italy)

\bibitem{35}
Ghosh S, Gadagkar V and Sood A K  2005 \emph{Chem. Phys. Lett.} \textbf{406} 10

\bibitem{36}
Lu Y, Goldsmith B R, Kybert N J and Johnson A T C 2010 \emph{Appl. Phys. Lett.} \textbf{97} 083107
\end{thebibliography}
\end{document}